\documentclass{elsart}
\usepackage{graphicx}
\usepackage{amssymb}

\begin{document}

\begin{frontmatter}
\title{Semantic learning in autonomously active recurrent neural networks}

\author{Claudius Gros and Gregor Kaczor}
\address{Institute for Theoretical Physics\\
         Goethe University Frankfurt\\
         60054 Frankfurt/Main, Germany}

\begin{abstract}
The human brain is autonomously active, being characterized by a 
self-sustained neural activity which would be present even in the 
absence of external sensory stimuli. Here we study the
interrelation between the self-sustained activity 
in autonomously active recurrent neural nets
and external sensory stimuli.

There is no a priori semantical relation between the influx 
of external stimuli and the patterns generated internally 
by the autonomous and ongoing brain dynamics. The question
then arises when and how are semantic correlations between
internal and external dynamical processes learned and built up?

We study this problem within the paradigm of transient state
dynamics for the neural activity in recurrent neural nets, 
i.e. for an autonomous neural activity characterized by 
an infinite time-series of transiently stable 
attractor states. We propose that external stimuli will be 
relevant during the sensitive periods, {\it viz} the transition period
between one transient state and the subsequent semi-stable attractor.
A diffusive learning signal is generated unsupervised whenever
the stimulus influences the internal dynamics qualitatively.

For testing we have presented to the model system stimuli 
corresponding to the bars and stripes problem. 
We found that the system performs a non-linear 
independent component analysis on its own,  
being continuously and autonomously active. 
This emergent cognitive capability results here 
from a general principle for the neural dynamics, the
competition between neural ensembles.
\end{abstract}

\begin{keyword}
recurrent neural networks\sep autonomous neural dynamics\sep 
transient state dynamics\sep emergent cognitive capabilities
%
\end{keyword}
\end{frontmatter}

\section{INTRODUCTION}

It is well known that the brain has a highly developed
and complex self-generated dynamical neural activity. 
We are therefore confronted with a dichotomy when
attempting to understand the overall functioning of
the brain or when designing an artificial cognitive
system: A highly developed cognitive system, such as
the brain \cite{fiser2004}, is influenced by sensory input 
but it is not driven directly by the input signals. The cognitive 
system needs however this sensory information vitally for 
adapting to a changing environment and survival.

In this context we then want to discuss two mutually 
interrelated questions:
\begin{itemize}
\item Can we formulate a meaningful paradigm for the self-sustained
      internal dynamics of an autonomous cognitive system?
\item How is the internal activity influenced by sensory signals,
      {\it viz} which are the principles for the respective learning processes?
\end{itemize}
We believe that these topics represent important
challenges for research in the field of recurrent neural networks and
the modeling of neural processes. From an experimental point
of view we note that an increasing flux of results from neurobiology 
supports the notion of quasi-stationary spontaneous neural activity 
in the cortex 
\cite{fiser2004,abeles95,ringach03,kenet03,damoiseaux06,Honey07}.
It is therefore reasonable to investigate the two 
questions formulated above
with the help of neural architectures centrally based 
on the notion of spontaneously generated transient states,
as we will do in the present investigation using appropriate
recurrent neural networks.

\subsection{Transient-state and competitive dynamics}

\begin{figure}[tb]
\centerline{
\includegraphics[width=0.85\textwidth]{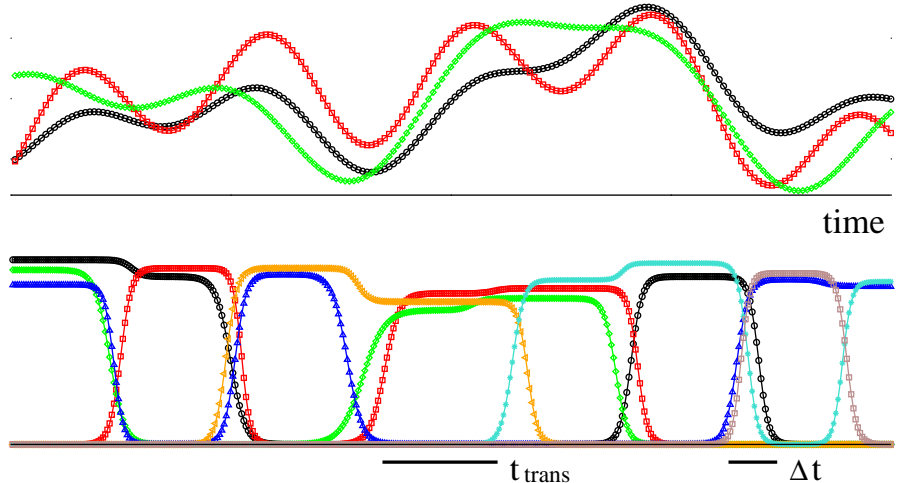}
           }
\caption{Schematic illustration of general or fluctuating
activity patterns (top) and of transient-state
dynamics (bottom), which is characterized by
typical time scales $t_{trans}$ and $\Delta t$
for the length of activity-plateau and of the
transient period respectively.}
\label{figure_trans_activity}
\end{figure}

Standard classification schemes of dynamical systems
are based on their long-time behavior, which may
be characterized, e.g., by periodic or 
chaotic trajectories \cite{grosBook08}.
The term `transient-state dynamics'
refers, on the other hand, 
to the type of activity occurring on
intermediate time scales, as illustrated in
Fig.~\ref{figure_trans_activity}. A time
series of semi-stable activity patterns,
also denoted transient attractors, is
characterized by two time scales.
The typical duration $t_{trans}$ of
the activity plateaus and the typical time 
$\Delta t$ needed to perform the transition from
one semi-stable state to the subsequent one.
The transient attractors turn into
stable attractors in the limit $t_{trans}/\Delta t\to\infty$.

Transient state dynamics is intrinsically 
competitive in nature. When the current transient
attractor turns unstable the subsequent transient
state is selected by a competitive process.
Transient-state dynamics is a form of `multi-winners-take-all'
process, with the winning coalition of dynamical
variables suppressing all other competing activities.

Humans can discern about 10-12 objects per second 
\cite{vanrullen2003} and it is therefore tempting
to identify the cognitive time scale of about 80-100ms
with the duration $t_{trans}$ of the transient-state
dynamics illustrated in Fig.~\ref{figure_trans_activity}. 
Interestingly, this time scale also coincides with the
typical duration \cite{kenet03} of the transiently active 
neural activity patterns observed in the cortex 
\cite{abeles95,ringach03,damoiseaux06,Honey07}. 

Several high-level functionalities have been proposed 
for the spontaneous neural brain dynamics.
Edelman and Tononi \cite{edelman00,edelman03} argue that
`critical reentrant events' constitute transient
conscious states in the human brain. These
`states-of-mind' are in their view semi-stable global activity states
of a continuously changing ensemble of neurons, the `dynamic core'.
This activity takes place in what Dehaene and Naccache \cite{dehaene03}
denote the `global workspace'. The global workspace serves, in the
view of Baars and Franklin \cite{baars03},
as an exchange platform for conscious experience 
and working memory. Crick and Koch \cite{crick03} 
and Koch \cite{koch04} have
suggested that the global workspace is made-up of
`essential nodes', i.e. ensembles of neurons responsible
for the explicit representation of particular aspects
of visual scenes or other sensory information.

\subsection{Autonomously active recurrent neural nets}

Traditional neural network architectures are not
continuously active on their own. Feedforward setups are
explicitly driven by external input \cite{haykin1994} 
and Hop\-field-type recurrent nets settle into a
given attractor after an initial period of 
transient activities \cite{hopfield1982}. The
possibilities of performing cognitive computation 
with autonomously active neural networks, the
route chosen by nature, are however investigated
increasingly \cite{grosKeynote2009}. In this 
context the time encoding of neural information,
one of the possible neural codes \cite{neuralCode1998},
has been studied in various contexts. Two network
architectures, the echo state network suitable
for rate-encoding neurons \cite{jaeger2004}, and
the liquid state machine suitable for spiking neurons
\cite{maass2002}, have been proposed to transiently
encode in time a given input for further linear
analysis by a subsequent perceptron. Both architectures,
the echo-state network and the liquid-state machine,
are examples of reservoir architectures with fading
memories, which however remain inactive in the absence 
of sensory input.

An example of a continuously active recurrent network
architecture is the winnerless competition based 
on stable heteroclinic cycles \cite{rabinovich2001}.
In this case the trajectory moves along heteroclines
from one saddle point to the next approaching a complex
limiting cycle. Close to the saddle points the dynamics
slows down leading to well defined transiently active
neural activity patterns.
\begin{figure}[tb]
\centerline{\hfill
\includegraphics[width=0.45\textwidth]{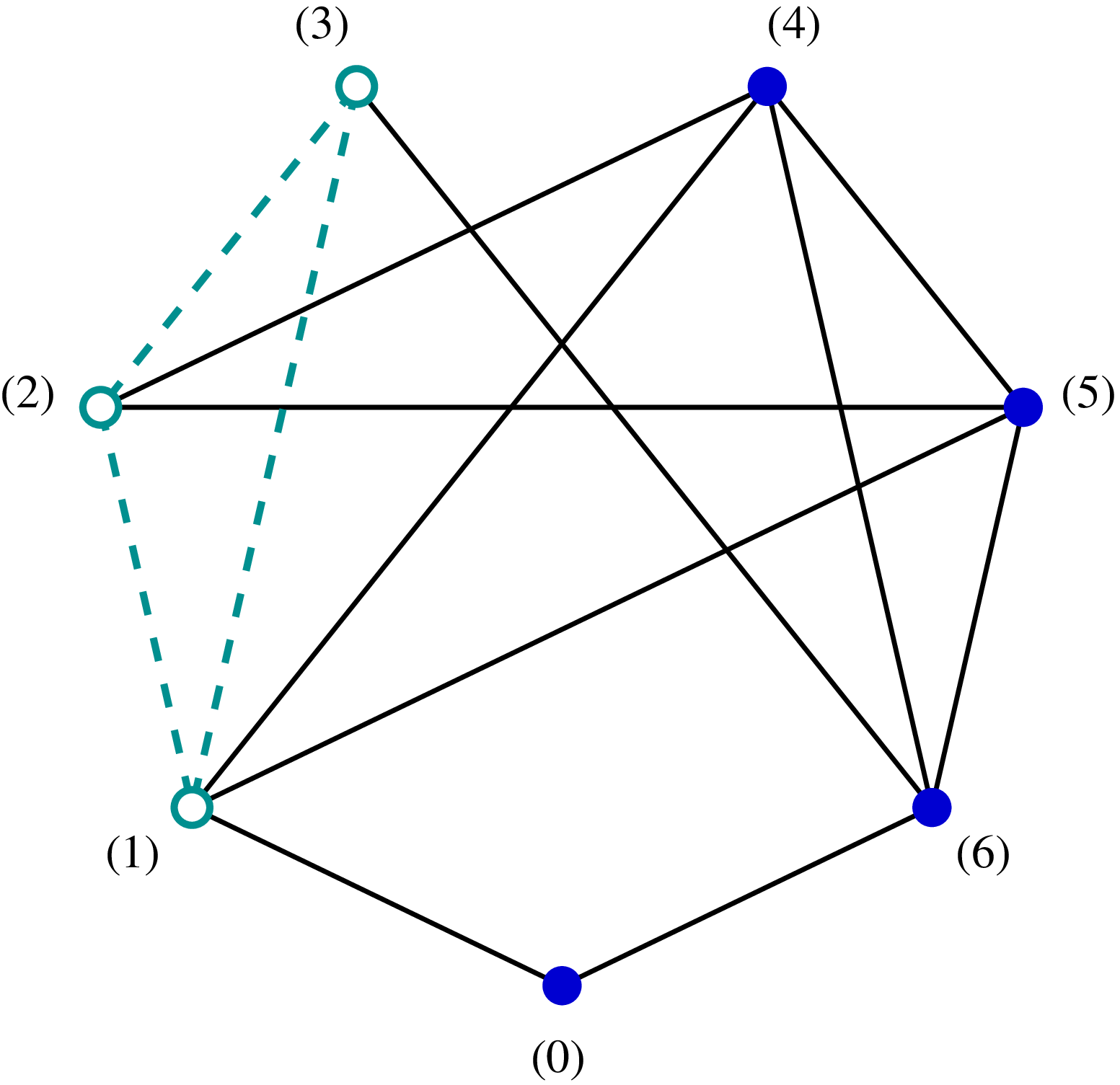} \hfill
\includegraphics[width=0.45\textwidth]{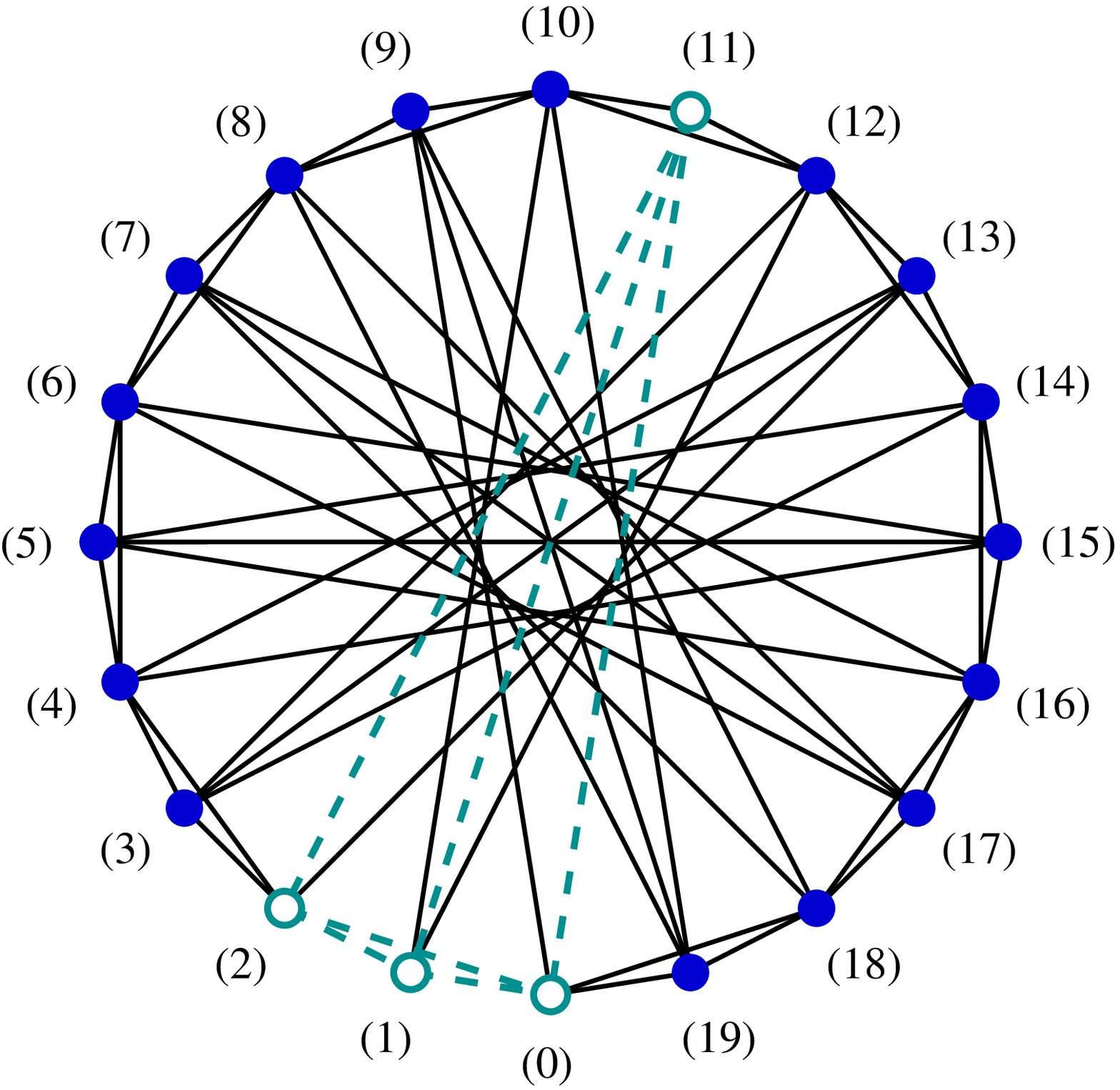}
            \hfill}
\caption{Illustration of clique encoding. Note that
cliques are fully connected subgraphs of maximal
size. The dashed lines are highlighted 
examples of specific cliques.
Left: A 7-site network with cliques (1,2,3),
(3,6), (1,2,4,5), (4,5,6), (0,6) and (0,1).
Right: A regular 20-site networks containing 10
four-site cliques. }
\label{figure_networks_cliques}
\end{figure}

\section{CLIQUE ENCODING IN RECURRENT NETWORKS}

In order to study the issues raised in the introduction,
the notion of autonomous neural activity and
its relation to the sensory input, we will consider
a specific model based on clique-encoding recurrent
nets. The emphasis will be on the discussion of 
the general properties and of the underlying challenges. We will
therefore present here an overview of the algorithmic
implementation, referring in part to the literature
for further details.

\subsection{Cliques, attractors and transient states}

Experimental evidence indicates that sparse neural 
coding is an important operating principle in the brain, 
as it minimizes energy consumption, maximizes
storage capacity and contributes to make 
information encoding spatially explicit \cite{sparseCoding2004}.
A powerful form of sparse coding is
multi-winners-take-all encoding in the form
of cliques. The term cliques stems from network theory 
and denotes subgraphs which are fully interconnected
\cite{grosBook08}, a few examples 
are given in Fig.~\ref{figure_networks_cliques}.
Cliques are fully interconnected subgraphs of
maximal size, in the sense that they are not part
of another fully interconnected subgraph containing
a larger number of vertices.

Clique encoding is an instance of sparse coding
with spatially overlapping memory states. The
use of clique encoding is in fact motivated by
experimental findings indicating a hierarchical
organization of overlapping neural clique assemblies 
for the real-time memory representation in the
hippocampus \cite{neuralCliques06}.
In the framework of a straightforward 
auto-associative neural network the cliques are
defined by the network of the excitatory connections,
which are shown as lines in 
Fig.~\ref{figure_networks_cliques}, in the presence
of an inhibitory background \cite{gros05,grosNJP07}.
In this setting all cliques correspond to
attractors of the network, {\it viz} to spatially
explicit and overlapping memory representations.

\begin{figure}[tb]
\centerline{
\includegraphics[width=0.55\textwidth]{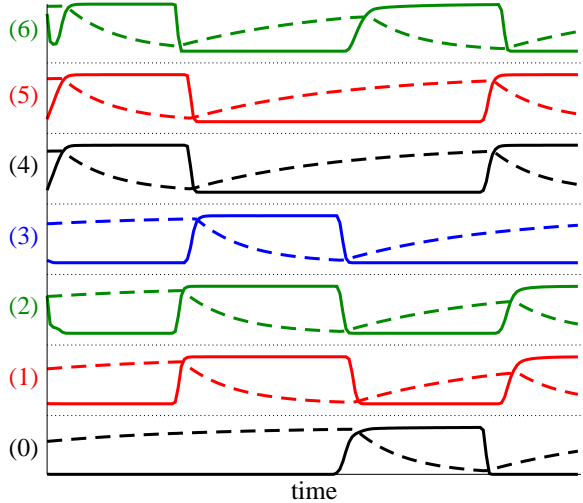}
           }
\caption{Transient state dynamics of the 7-site
network illustrated in Fig.~\ref{figure_networks_cliques}.
Shown are, vertically displaced, the time developments 
of the respective neural activities (solid lines) 
and of the neural reservoirs (dashed lines). The time series of
spontaneously generated transient states are the cliques
$(4,5,6)\to(1,2,3)\to(0,6) \to(1,2,4,5)$ (from the left
to the right). }
\label{figure_AI7}
\end{figure}

One can transform the attractor network
with clique encoding into a continuously active
transient-state network, by introducing a
reservoir variable for every neuron.
In this setting the reservoir of a neuron is
depleted whenever the neuron is active and refilled whenever
the neuron is inactive. Via a suitable local coupling between
the individual neural activity and reservoir variables
a well defined and stable transient state dynamics
is obtained \cite{gros05,grosNJP07}. When a given
clique becomes a winning coalition, the reservoirs
of its constituting sites are depleted over time.
When fully depleted the winning coalition becomes
unstable and the subsequent winning coalition is
activated through a competitive associative process,
leading to an ever ongoing associative thought process.
The resulting network architecture is a dense and
homogeneous associative network (dHan) \cite{gros05}. 
An illustrative result of a numerical simulation
is given in Fig.~\ref{figure_AI7}.

For the isolated system, not coupled to any sensory input,
this associative thought process has no semantic
content, as the transient attractors, the cliques,
have none. The semantic content can be acquired only
by coupling to a sensory input and by the generation
of correlation between the transient attractors and 
patterns extracted from the input data stream.

\subsection{Competitive dynamics and sensitive periods}

\begin{figure}[tb]
\centerline{
\includegraphics[width=0.85\textwidth]{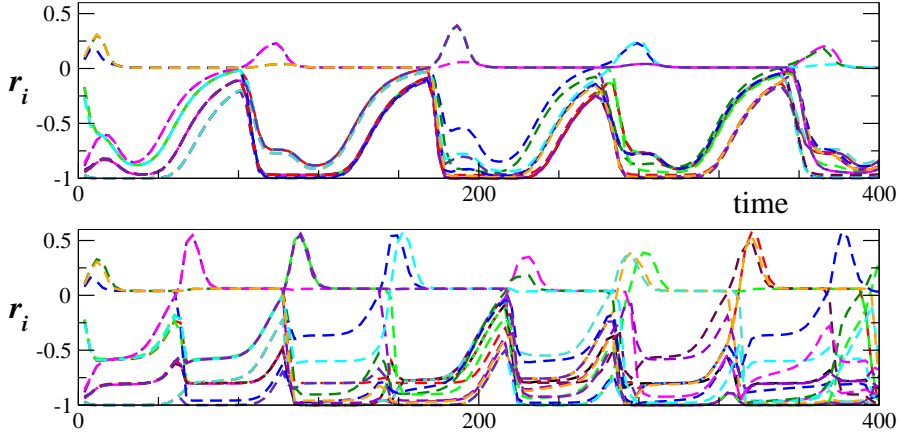}
           }
\caption{The growth rates $r_i(t)$ generating
an internal transient state-dynamics via 
Eq.~(\ref{eq_x_dot}). The two examples differ in the
functional dependence of the $r_i(t)$ on
the $x_j(t)$. The top graph corresponds to
a system having sensitive periods, the bottom
graph to a system without distinctive sensitive
periods.}
\label{figure_sensitive_periods}
\end{figure}

To be definite, we utilize a continuous-time formulation 
for the dHan architecture, with rate-encoding neurons, 
characterized by normalized activity levels
$x_i\in[0,1]$. One can then define, via
\begin{equation}
{d\over dt} x_i \ =\ 
\left\{
\begin{array}{rcl}
(1-x_i)\,r_i &\qquad& (r_i>0) \\
x_i\, r_i &\qquad& (r_i<0) 
\end{array}
\right.  
\label{eq_x_dot}
\end{equation}
the respective growth rates $r_i$.
Representative time series of growth rates $r_i$
are illustrated in Fig.~\ref{figure_sensitive_periods}.
When $r_i>0$, the respective neural 
activity $x_i$ increases, approaching rapidly the
upper bound; when
$r_i<0$, it decays to zero. The model is
specified \cite{gros05,grosNJP07},
by providing the functional dependence 
of the growth rates with respect
to the set of activity-levels $\{x_j\}$ of
all sites and on the synaptic weights, as 
usual for recurrent or auto-associative networks.

During the transition periods many, if not all,
neurons will enter the competition to become
a member of the new winning coalition. The
competition is especially pronounced whenever
most of the growth rates $r_i$ are small in magnitude,
with no subset of  growth rates dominating over
all the others. Whether this does or does not happen 
depends on the specifics of the model setup.
In Fig.~\ref{figure_sensitive_periods},
two cases are illustrated.
In the first case (lower graph) the competition for the next 
winning coalition is restricted to a subset of neurons, 
in the second case (upper graph) the 
competition is network-wide.
When most neurons participate in the competition 
process for a new winning
coalition the model will have `sensitive periods'
during the transition times and it will be able
to react to eventual external signals.

\subsection{Sensitive periods and learning}

So far we have discussed in general terms the properties
of isolated models exhibiting a self-sustained dynamical behavior
in terms of a never-ending time series of semi-stable
transient states, as illustrated
in Figs.~\ref{figure_AI7}
and \ref{figure_sensitive_periods},
using the dHan architecture with
continuous-time and rate-encoding neurons.

\begin{figure}[tb]
\centerline{
\includegraphics[width=0.75\textwidth]{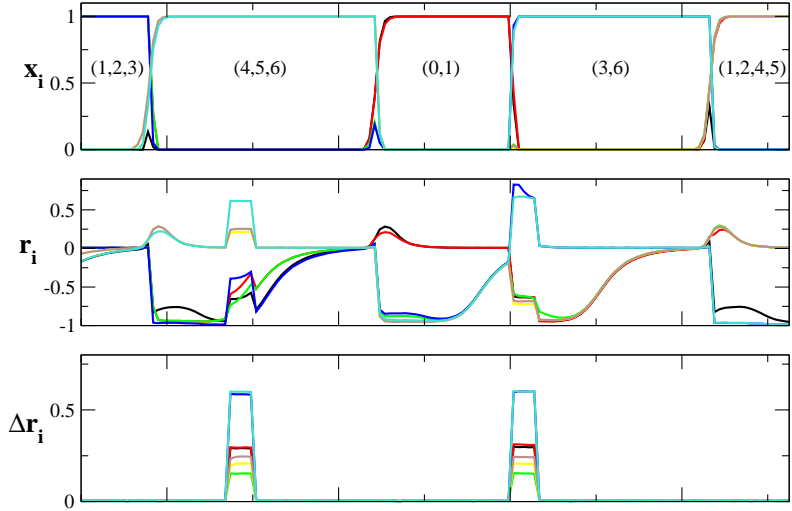}
           }
\caption{Results for the activities $x_i(t)$, the
total growth rates $r_i(t)$ and the input signals
$\Delta r_i(t)$ from a simulation of the 7-site system
shown in Fig.~\ref{figure_networks_cliques}
(color coding). The time series of winning coalitions 
is given. The first input signal does not change
the composition of the winning coalition, whereas the
second does.}
\label{figure_ARB_7}
\end{figure}

The importance of sensitive periods comes in 
when the network exhibiting trans\-ient-state
dynamics is coupled to a stream of sensory 
input signals. It is reasonable to assume, 
that external input signals will contribute 
to the growth rates $r_i$ via
\begin{equation}
r_i \ \equiv\  r_i^{dHan}\, +\, 
\Delta r_i\,\Big(\,1-\Theta(r_i^{dHan})\Theta(-\Delta r_i)\,\Big)~.
\label{eq_delta_r}
\end{equation}
Here the $\Delta r_i$ encode the influence of the input
signals and we have denoted now with $r_i^{dHan}$
the contribution to the growth rate a neuron in the dHan
layer receives from the other dHan neurons. 
The factor $(1-\Theta(r_i^{dHan})\Theta(-\Delta r_i))$
in Eq.~(\ref{eq_delta_r})
ensures that the input signal does not deactivate
the current winning coalition as we will discuss
further below. Let us here assume for a moment, 
as an illustration, that the input signals are 
suitably normalized, such that
\begin{equation}
\Delta r_i \ \simeq\  \left\{
\begin{array}{rcl}
0.5 &\qquad& \mbox{(active input)} \\
0 &\qquad& \mbox{(inactive input)} 
\end{array}
\right.  ,
\label{eq_delta_r_magnitude}
\end{equation}
in order of magnitude. For the simulations presented
further below a qualitatively similar optimization will
occur homeostatically. For the transient states
the $r_i^{dHan}\approx -1$ for all sites not
forming part of the winning coalition and the 
input signal $\Delta r_i$ will therefore not 
destroy the transient state, compare 
Figs.~\ref{figure_sensitive_periods}
and \ref{figure_ARB_7}.
With the normalization
given by Eq.~(\ref{eq_delta_r_magnitude}) the 
total growth rate $\sim(\Delta r_i-1)$ will remain 
negative for all inactive sites and the sensory
input will not be able to destroy the current
winning coalition. The input signal will however
enter the competition for the next
winning coalition during a sensitive period, 
providing an additional boost for the 
respective neurons. 

This situation is exemplified in
Fig.~\ref{figure_ARB_7}, where we present
simulation results for the 7-site system 
shown in Fig.~\ref{figure_networks_cliques},
subject to two sensory 
inputs $\Delta r_i(t)$. The self-generated
time series of winning coalitions is not
redirected for the first sensory input.
The second stimulus overlaps with a sensory
period and its strongest components determine
the new winning coalitions.
The simulation results presented in
Fig.~\ref{figure_ARB_7} therefore demonstrate
the existence of well defined time-windows suitable 
for the learning of correlations between the 
input signal and the intrinsic dynamical activity. 
The time windows, or sensitive periods, are
present during and shortly after a transition 
from one winning coalition to the subsequent. 
A possible concrete implementation for this type of
learning algorithm will be given further below.

Let us now come to the factor 
$(1-\Theta(r_i^{dHan})\Theta(-\Delta r_i))$
in Eq.~(\ref{eq_delta_r}), containing the
Heaviside-step functions $\Theta(x)$. For 
vertices $i$ of the current winning coalition
the intra dHan-layer growth rates are positive, 
$r_i^{dHan}>0$. Therefore, the above factor 
ensures, that a suppressive $\Delta r_i<0$
has no effect on the members of the current
winning coalition. The contribution $\Delta r_i$
from the input may therefore alter the balance
in the competition for the next winning coalition
during the sensitive periods, but not suppress
the current active clique.

Let us note, that the setup discussed here
allows the system also to react to an 
occasional strong excitatory input signal 
having $\Delta r_i>1$. Such a strong signal
would suppress the current transient state 
altogether and impose itself. This possibility of 
rare strong input signals is evidently important for
animals and would be, presumably, also helpful
for an artificial cognitive system.

\subsection{Diffusive learning signals}

Let us return to the central problem inherent
to all systems reacting to input signals and having
at the same time a non-trivial intrinsic dynamical 
activity. Namely, when should learning occur, i.e.\
when should a distinct neuron become more
sensitive to a specific input pattern and
when should it suppress its sensibility to
a sensory signal.

The framework of competitive dynamics developed
above allows for a straightforward solution
of this central issue: Learning should occur
exclusively when the input signal makes
a qualitative difference, {\it viz} when the
input signal deviates the transient-state
process. For illustration let us assume that
the series of winning coalitions is
$$
(4,5,6)\ {[a]\atop\longrightarrow}\ (0,1) 
     \ {[a]\atop\longrightarrow}\ (1,2,4,5)~,
$$
where the index $[a]$ indicates that the
transition is driven by the autonomous 
internal dynamics and that the series of
winning coalitions take the form
$$
(4,5,6)\ {[a]\atop\longrightarrow}\ (0,1) 
     \ {[s]\atop\longrightarrow}\ (3,6)~,
$$
in the presence of a sensory signal $[s]$,
as it is the case for the data presented in
Fig.~\ref{figure_ARB_7}.
Note, that a background of weak or noisy sensory input 
could be present in the first case, but learning should 
nevertheless occur only in the second case.
A reliable distinction between these two cases
can be achieved via a
suitable diffusive learning signal\footnote{The 
name `diffusive learning signal' 
\cite{grosBook08} stems from the fact, 
that many neuromodulators are
released in the brain in the intercellular 
medium and then diffuse physically to the surrounding
neurons, influencing the behavior of large
neural assemblies.} $S(t)$. It is activated
whenever any of the input contributions $\Delta r_i$
changes the sign of the respective growth
rates during the sensitive periods,
\begin{equation}
{d\over dt} S \ \to\  \left\{
\begin{array}{rcl}
\phantom{-}\Gamma_{diff}^+ &\qquad& (r_i>0) 
             \mbox{\ and\ } (r_i^{dHan}<0)\\
-\Gamma_{diff}^- &\qquad& \mbox{otherwise} 
\end{array}
\right. ,
\label{eq_dot_S}
\end{equation}
{\it viz} when it makes a qualitative
difference. Let us remember that the $r_i^{dHan}$ 
are the internal contributions to the growth rate,
i.e.\ the input a dHan neuron receives via
recurrent connections from the other dHan neurons.
The diffusive learning signal $S$ is therefore
increasing in strength only when a neuron is 
activated externally, but not when activated
internally, with the $\Gamma_{diff}^{\pm}>0$ denoting
the respective growth and decay rates.
The diffusive learning signal $S(t)$ is a 
global signal and a sum $\sum_i$ over all dynamical
variables is therefore implicit on the
right-hand side of Eq.~(\ref{eq_dot_S}).

\subsection{The role of attention}

The general procedure for the learning of
correlation between external signals and
intrinsic dynamical states for a cognitive
system presented here does not rule out
other mechanisms. Here we concentrate on the
learning algorithm which occurs automatically,
one could say sub-consciously. Active
attention focusing, which is well known in
the brain to potentially shut off a sensory 
input pathway, or to enhance sensibility
to it, may very well work in parallel to
the continuously ongoing mechanism investigated
here.

We note, however, that the associative thought process
within the dHan carries with it a dynamical 
attention field \cite{gros05}. Neurons receiving both
positive and negative contributions from the
winning coalition will need smaller sensory
input signals in order to be activated than neurons
receiving only negative contributions.
To put it colloquially: When thinking of the color
blue it is easier to spot a blue car in the traffic
than a white one.

\begin{figure}[t]
\centerline{
\includegraphics*[width=0.85\textwidth]{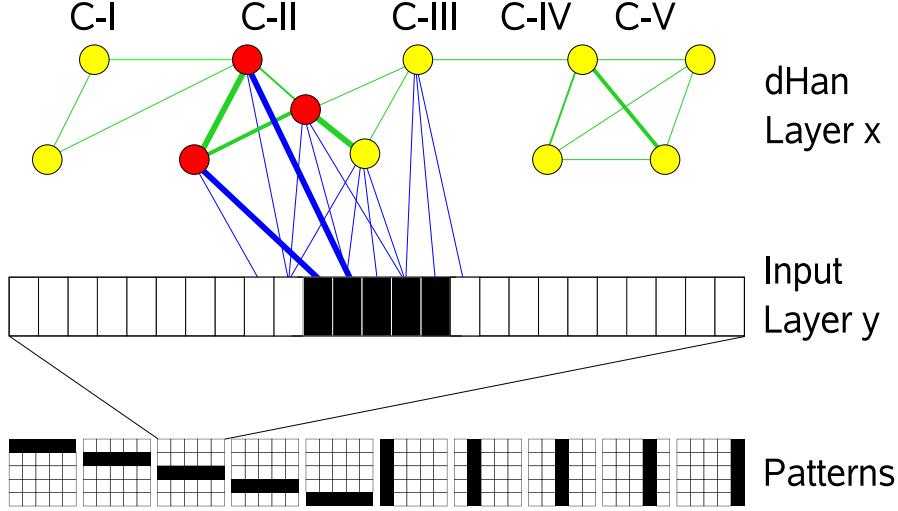}
           }
\caption{Schematic representation of the information flow from a raw pattern 
(bottom) via the input layer (middle) to the dHan layer (top). 
The synaptic strengths $v_{ij}$ connecting the input with the
dHan layer are adapted during the learning process (illustrated 
selectively by respective thick/thin blue lines in the graph). 
The dHan layer consists of active and inactive neurons 
(red/yellow circles) connected by intra-layer synaptic weights.
The topology shows five cliques (denoted C-I to C-V in the graph)
of which C-II is active, as emphasized by the red-color neurons.
        }
\label{fig_twoLayers}
\end{figure}

\section{COMPETIVE LEARNING}

So far we have described, in general terms, the system
we are investigating. It has sensitive periods during
the transition periods of the continuously ongoing
transient-state process, with learning of input
signals regulated by a diffusive learning signal.
The main two components are therefore the dHan layer
and the input layer, as illustrated
in Fig.~\ref{fig_twoLayers}. 

\subsection{Input data-stream analysis}

The input signal acts via Eq.~(\ref{eq_delta_r}) on
the dHan layer, with the contribution $\Delta r_i$
to the growth rate of the dHan neuron $i$
given by
\begin{equation}
\Delta r_i = \sum_{j} v_{ij} y_j,
\qquad\quad
\Delta s_i = \sum_{j} v_{ij} (1-y_j),
\label{eq_Delta_r_v_pq}
\end{equation}
where we have denoted with $y_j\in[0,1]$ the activity-levels
of the neurons in the input layer.  For subsequent
use we have defined in Eq.~(\ref{eq_Delta_r_v_pq}) an
auxiliary variable $\Delta s_i$, which quantifies
the influence of inactive input-neurons. The task
is now to find a suitable learning algorithm which
extracts relevant information from the input-data
stream by mapping distinct input-patterns onto
selected winning coalitions of the dHan layer.
This setup is typical for an independent
component analysis \cite{hyvarinen00}.

The multi-winners-take-all dynamics in the dHan
module implies individual neural activities to
be close to 0/1 during the transient states 
and we can therefore define three types of
inter-layer links $v_{ij}$ (see
Fig.~\ref{fig_twoLayers}):

\begin{itemize}
\item \underline{active} ({\it `act'})\\
      Links connecting active input neurons
      with the winning coalition of the
      dHan module. 
\item \underline{orthogonal} ({\it `orth'})\\
      Links connecting inactive input neurons
      with the winning coalition of the
      dHan module.
\item \underline{inactive} ({\it `ina'})\\
      Links connecting active input neurons
      with inactive neurons of the
      dHan module.
\end{itemize}
The orthogonal links
take their name from the circumstance that
the receptive fields of the winning coalition
of the target layer need to orthogonalize to
all input-patters differing from the present one.
Note that it is not the receptive field of individual
dHan neurons which is relevant, but rather 
the cumulative receptive field of a given winning
coalition.

We can then formulate three simple rules for the 
respective link-plasticity. Whenever
the new winning coalition in the dHan layer
is activated by the input layer, {\it viz}
whenever there is a substantial 
diffusive learning signal, 
i.e.\ when $S_{diff}$ exceeds a certain
threshold $S_{diff}^c$, the following
optimization procedures should take
place:
\begin{itemize}
\item \underline{active links}\\
      The sum over active afferent links should take
      a large but finite value $r_v^{act}$,
$$
\sum_{j} v_{ij}\, y_j \bigg|_{x_i\,{\rm active}}
 \ \to\ r_v^{act}~.
$$
\item \underline{orthogonal links}\\
      The sum over orthogonal afferent links should take
      a small value $s_v^{orth}$,
$$
\sum_{j} v_{ij}\,(1-y_j) \bigg|_{x_i\,{\rm active}}
 \ \to\ s_v^{orth}~.
$$
\item \underline{inactive links}\\
      The sum over inactive links should take
      a small but non-vanishing value $r_v^{ina}$,
$$
\sum_{j} v_{ij}\, y_j \bigg|_{x_i\,{\rm inactive}}
 \ \to\ r_v^{ina}~.
$$
\end{itemize}
The $r_v^{act}$, $r_v^{ina}$ and
$s_v^{orth}$ are the target values for the
respective optimization processes.
In order to implement these three rules 
we define three corresponding contributions
to the link plasticities:
\begin{equation}
\begin{array}{rcl}
c_i^{act} & =& \Gamma_v^{act}\,\Theta(x_i-x_v^{act})
               \,{\rm Sign}(r_v^{act}-\Delta r_i) \\
c_i^{orth} & =& \Gamma_v^{orth}\,\Theta(x_i-x_v^{act})
               \,{\rm Sign}(s_v^{orth}-\Delta s_i) \\
c_i^{ina} & =& \Gamma_v^{ina}\,\Theta(x_v^{ina}-x_i)
               \,{\rm Sign}(r_v^{ina}-\Delta r_i) 
\end{array}
\label{eq_back_contributions}
\end{equation}
where the inputs $\Delta r_i$ and
$\Delta s_i$ to the dHan layer
are defined by Eq.~(\ref{eq_Delta_r_v_pq}). For the
sign-function ${\rm Sign}(x)=\pm1$ is valid,
for $x>0$ and $x<0$ respectively, 
$\Theta(x)$ denotes the Heaviside-step
function. In Eq.~(\ref{eq_back_contributions}) the
$\Gamma_v^{act}$, $\Gamma_v^{orth}$ and $\Gamma_v^{ina}$
are suitable optimization rates and
the $x_v^{act}$ and $x_v^{ina}$ the 
activity levels defining active and
inactive dHan neurons respectively. 
A suitable set of parameters, which has been used
for the numerical simulations, 
is given in Table~\ref{tab_par_back}.

\begin{table}[b]
\caption{The set of parameters entering the
time-evolution equations of the
links connecting the input to the dHan layer, 
with $\Gamma_{diff}^+=4.0$ and $\Gamma_{diff}^-=0.15$,
used in the actual simulations.
        }
\begin{center}
\begin{tabular}{c|c|c|c}
\hline \hline
$ \Gamma_v^{act} $ \ $ \Gamma_v^{orth} $ \ 
$ \Gamma_v^{ina} $ & $ r_v^{act}       $ \ 
$ s_v^{orth}     $ \ $ r_v^{ina}       $ & 
$ x_v^{act}      $ \ $ x_v^{ina}       $ &
$ S_{diff}^c     $ 
\\
\hline
0.002 \ 0.001 \ 0.001 & 0.8 \ \ \ 0.2 \ \ \ 0.2 & 0.4 \ \ \ 0.2 & 0.25 \\
\hline \hline
\end{tabular}
\end{center}
\label{tab_par_back}
\end{table}

Using these definitions, the link plasticity
may be written as
\begin{equation} 
\label{eq_vdot}
{d\over dt} v_{ij} \ =\ \Theta(S_{diff}-S_{diff}^c)\,
\Big[\, c_i^{act}\,y_j 
 + c_i^{orth}\,\left(1-y_j\right) 
 + c_i^{ina}\,y_j\,\Big]~, 
\nonumber 
\end{equation}
where $S_{diff}^c$ is an appropriate threshold for
the diffusive learning signal.
The inter-layer links $v_{ij}$ cease
to be modified whenever the total input
is optimal, {\it viz} when no more `mistakes` 
are made \cite{chialvo99}.

We note, that a given interlayer-link
$v_{ij}$ is in general subject to
competitive optimization from the three
processes (act/orth/ina). Averaging
would occur if the respective learning
rates $\Gamma_v^{act}$/$\Gamma_v^{orth}$/$\Gamma_v^{ina}$
would be of the same order of magnitude.
It is therefore necessary, that
$ \Gamma_v^{act}\gg\Gamma_v^{orth}$ and
$\Gamma_v^{act}\gg\Gamma_v^{ina}$.

\begin{figure}[tb]
\centerline{
\includegraphics[width=0.85\textwidth]{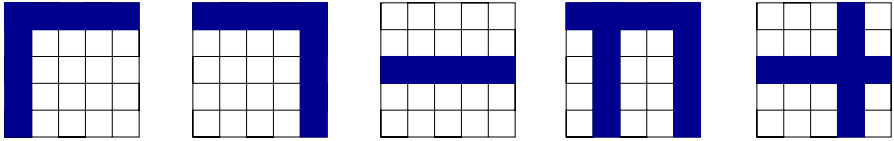}
           }
\caption{Examples of typical input patterns for a
$5\times5$ bars problem with a probability $p=0.1$
for the occurrence of the individual horizontal or
vertical bars. The problem is non-linear since the
pattern intensity is not enhanced when an elementary
horizontal and vertical bar overlap each other.}
\label{figure_bars_pattern}
\end{figure}

\subsection{Homeostatic normalization}

It is desirable that the interlayer connections 
$v_{ij}$ neither grow unbounded with
time (runaway-effect) nor disappear into 
irrelevance. Suitable normalization procedures
are therefore normally included explicitly into the 
respective neural learning rules and are present implicitly
in Eqs.~(\ref{eq_back_contributions}) and
Eq.~(\ref{eq_vdot}). 

The strength of the input-signal is optimized
by Eq.~(\ref{eq_vdot}) both for active as well as
for inactive dHan neurons, a property referred to as
fan-in normalization. Eqs.~(\ref{eq_back_contributions}) 
and (\ref{eq_vdot}) also regulate the overall
strength of inter-layer links emanating from
a given input layer neuron, a property called
fan-out normalization. 

Next we note, that the time scales for the intrinsic autonomous
dynamics in the dHan layer and for the input signal could in
principle differ substantially. Potential
interference problems can be avoided when
learning is switched-on very fast. In this 
case the activation and decay rates
$\Gamma_{diff}^\pm$ for the diffusive
learning signal are large and the
corresponding characteristic time scales
$1/\Gamma_{diff}^\pm$ are smaller
than both the typical time scales of the
input and of the self-sustained dHan dynamics.

\section{THE BARS PROBLEM}

A cognitive system needs to extract autonomously
meaningful information about its environment from its
sensory input data stream via signal separation
and feature extraction. The identification of recurrently
appearing patterns, i.e.\ of objects, in the background of fluctuation
and of combinations of distinct and noisy patterns, constitutes 
a core demand in this context. This is the domain of
the independent component analysis \cite{hyvarinen00} 
and blind source separation \cite{blindSource},
which seeks to find distinct representations of statistically
independent input patterns.
 
In order to test our system made-up by an input layer coupled
to a dHan layer, as illustrated in Fig.~\ref{fig_twoLayers},
we have selected the bars problem \cite{barsProblem,triesch07}.
The bars problem constitutes a standard non-linear reference 
task for the feature extraction via a non-linear independent component 
analysis for an $L\times L$ input layer. Basic patterns 
are the $L$ vertical and $L$ horizontal bars. The 
individual input patterns are made-up of a non-linear superposition
of the $2L$ basic bars, containing with probability $p=0.1$ 
any one of them, as illustrated in Fig.~\ref{figure_bars_pattern}.

\begin{figure}[tb]
\centerline{
\includegraphics[width=0.95\textwidth]{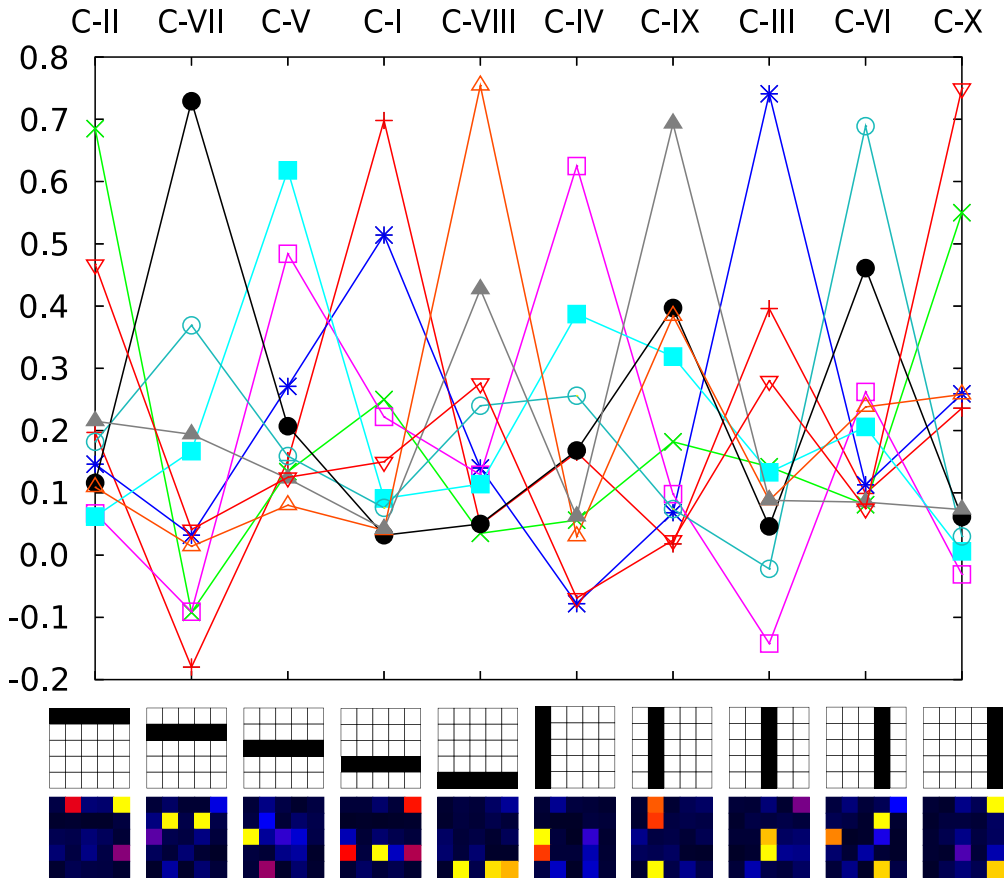}
           }
\caption{For the $5\times5$ bars problem
the response, as defined by Eq.~(\ref{eq_clique_rec_patt}),
for the 10 winning coalitions in the dHan layer
(compare Fig.~\ref{figure_networks_cliques} and 
         Fig.~\ref{fig_twoLayers})
to the ten reference patterns, {\it viz} the 5 horizontal bars 
and the 5 vertical bars of the $5\times5$ input field.
In the top row the numbering of the cliques C-I,..,C-X having
the maximal response to the respective reference patterns 
is given. In the bottom row, below each of the 
10 black/white reference patterns, the
receptive fields, Eq.~(\ref{eq_clique_recp_fields}),
for the winning coalitions C-I,..,C-X given in the top
row are given color-coded, with black/blue/red/yellow
coding synaptic strengths of increasing intensities.
}
\label{figure_cRP_graph}
\end{figure}

\subsection{Simulations and setup}

For the simulations we presented to the system about
$N_{patt}\approx 5\times 10^3$ randomly generated $5\times 5$ input 
patterns of the type shown in Fig.~\ref{figure_bars_pattern}.
The bars pattern are black/white with the $y_i=1/0$ for
active/inactive sites, irrespectively of possible overlaps
of vertical and horizontal bars. The individual
patterns lasted $T_{patt}=20$ with about $T_{inter}=100$
for the time between two successive input signals. These
time scales are to be compared with the time scale of
the autonomous dHan dynamics illustrated in
the Figs.~\ref{figure_sensitive_periods} and
\ref{figure_ARB_7}, for which the typical stability-period
for a transient state is about $t_{trans}\approx 70$.
We also note that there is no active training for the
system. The associative thought process continuous
in the dHan layer, at no time are the neural
activities reset and the system restarted. All
that happens is that the ongoing associative thought 
process is influenced from time to time by the
input layer and that then the synaptic strengths
$v_{ij}$ connecting the input layer to the dHan layer
are modified via Eq.~(\ref{eq_vdot}).

The results for the simulation are presented in
Fig.~\ref{figure_cRP_graph}.
For the geometry of the dHan network we used
a regular 20-site star containing 10
cliques, with every clique being composed
of four neurons, see Fig.~\ref{figure_networks_cliques}.
In Fig.~\ref{figure_cRP_graph} we present the  response
\begin{equation}
R(\alpha,\beta) \ =\ {1\over S(C_\alpha)} 
\sum_{i\in C_\alpha,j} v_{ij} y_j^{\beta},
\qquad 
\begin{array}{rcl}
&&\alpha=1,..,10 \\
&&\beta = 1,..,10
\end{array}
\label{eq_clique_rec_patt}
\end{equation}
of the 10 cliques $C_\alpha$ in the dHan
layer to the 10 basic input patterns
$\{y_j^{\beta},j=1,..,25\}$, the isolated bars. Here the
$C_\alpha\in\big\{\mbox{C-I,..,C-X}\big\}$ denotes the set 
of sites of the winning-coalition $\alpha$ and
$S(C_\alpha)$ its size, here $S(C_\alpha)\equiv4$.
The response $R(\alpha,\beta)$ is equivalent to
the clique averaged afferent synaptic signals 
$\Delta r_i$, compare Eq.~(\ref{eq_Delta_r_v_pq}), 
in the presence of an elementary bar in the sensory 
input field.

\subsection{Semantic learning}

The individual potential winning coalitions, {\it viz} the
cliques, have acquired in the course of the 
simulation, via the
learning rule Eq.~(\ref{eq_vdot}),
distinct susceptibilities to the 10 bars,
compare Fig.~\ref{figure_cRP_graph}. At the
start of the simulation the winning coalitions
were just given by properties of the network typology,
{\it viz} by the cliques, having no explicit semantic
significance. The susceptibilities to
the individual bars, which the cliques have
acquired via the competition of the
internal dHan dynamics with the sensory
data input stream, can then be interpreted
as a semantic assignment. The internal associative
thought process of the dHan layer therefore becomes
semantically meaningful via the coupling to
the environment, corresponding to a sequence of
horizontal and vertical bars. This learning paradigm
is compatible with multi-electrode array studies of
the visual cortex of developing ferrets \cite{fiser2004},
which indicate that the ongoing cortical dynamics
is void of semantic content immediately after birth,
acquiring semantic content however during the adolescence.

\subsection{Competitive learning}

The winning coalitions of the dHan layer are overlapping 
and every link $v_{ij}$ targets in general more than one
potential winning coalition in the dHan layer. This 
feature contrasts with the `single-winner-takes-all' setup, 
normally used for standard neural algorithms
performing an independent component analysis
\cite{hyvarinen00}, for which the target neurons 
are physically separated.
For the regular 20-site network used in the
simulation every dHan neuron appertains to exactly
two cliques, compare Fig.~\ref{figure_networks_cliques}.
The unsupervised learning procedure, Eq.~(\ref{eq_vdot}), 
involves therefore a competition
between the contribution $c_i^{act}$,
$c_î^{orth}$ and $c_î^{ina}$, as given
by Eq.~(\ref{eq_back_contributions}). 
For the simulations we used a set of parameters,
see Table~\ref{tab_par_back}, for which
the contribution to $c_i^{act}$ is
adapted at a much higher rate than
the contributions to $c_î^{orth}$ and $c_î^{ina}$.
The responses $R(\alpha,\beta)$ of the
winning coalitions are therefore close to,
but somewhat below, the optimal value
$r_v^{act}=0.8$ used for the simulations,
compare Fig.~\ref{figure_cRP_graph}. 
The target value $r_v^{act}$ will not be
reached even for extended simulations, 
due to the competition with the other
optimization procedures, namely $c_î^{orth}$ 
and $c_î^{ina}$, 
compare Eq.~(\ref{eq_back_contributions}).

\begin{figure}[tb]
\centerline{
\includegraphics[width=0.85\textwidth]{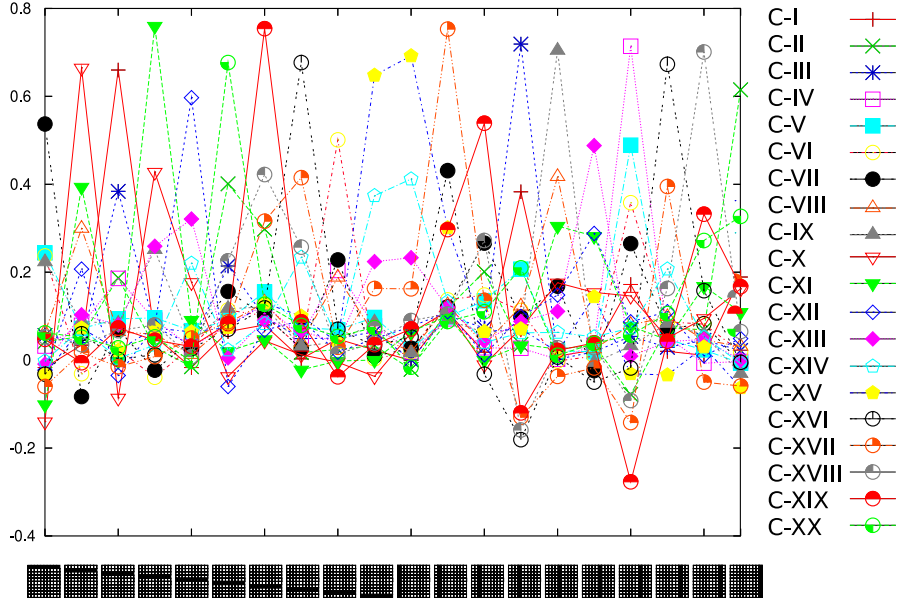}
           }
\caption{For the $10\times10$ bars problem the
response, as defined by Eq.~(\ref{eq_clique_rec_patt}),
for the 20 winning coalitions C-I,..,C-XX
in the dHan layer
(compare Fig.~\ref{figure_networks_cliques} and 
         Fig.~\ref{fig_twoLayers})
to the twenty reference patterns, {\it viz} the 10 horizontal bars 
and the 10 vertical bars of the $10\times10$ input field.
         }
 \label{figure_cRP_100}
 \end{figure}

\subsection{Receptive fields}

The averaged receptive fields
\begin{equation}
F(\alpha,j)\ =\ 
{1\over S(C_\alpha)} \sum_{i\in C_\alpha} v_{ij},
\qquad \alpha=1,..,10,
\label{eq_clique_recp_fields}
\end{equation}
of the $\alpha=1,...,10$ cliques in the dHan layer with 
respect to the $j=1,...,25$ input neurons are also
presented in Fig.~\ref{figure_cRP_graph}. 
The inter-layer synaptic weights
$v_{ij}$ can be both positive and negative and
the orthogonalization procedure, Eq.~(\ref{eq_back_contributions}),
results in complex receptive fields. 
The time evolution equations for the 
inter-layer synaptic strengths (\ref{eq_vdot})
are optimizing, but not maximizing, the response
of the winning coalition to a given input signal.
The receptive fields retain consequently a
certain scatter, since the optimization
via Eq.~(\ref{eq_vdot}) ceases whenever
a satisfactory signal separation has been obtained.
This behavior is consistent with the 
`learning by mistakes' paradigm \cite{chialvo99},
which states that a cognitive system needs to
learn in general only when committing a mistake.

\begin{figure}[tb]
\centerline{
\includegraphics[width=0.85\textwidth]{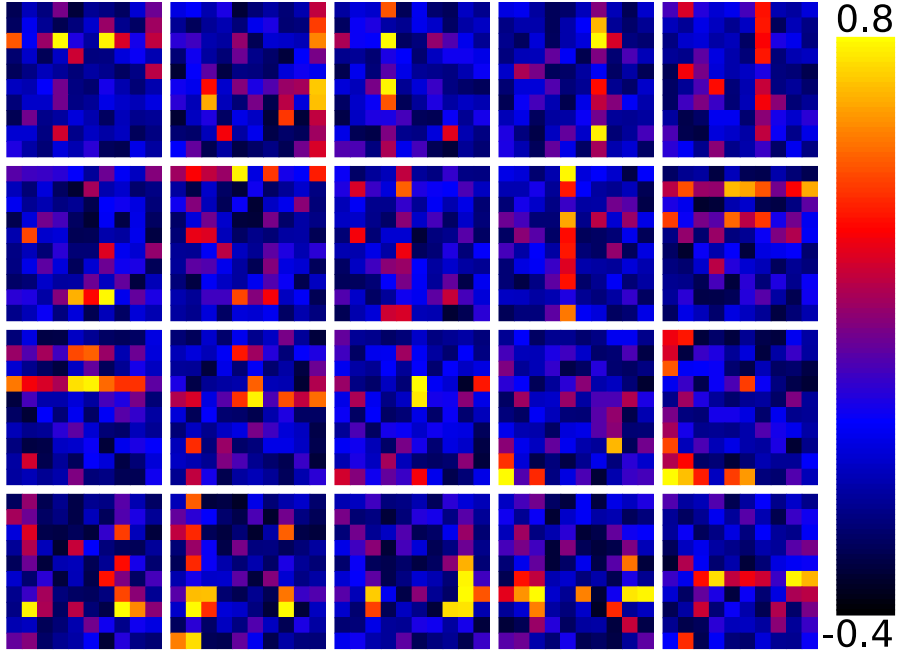}
           }
\caption{
For the $10\times10$ bars problem the color-coded 
receptive fields, Eq.~(\ref{eq_clique_recp_fields}),
for the 20 cliques C-I,..,C-XX of the dHan layer, 
compare Fig.~\ref{figure_cRP_100},
with black/blue/red/yellow coding synaptic 
strengths of increasing intensities. 
         }
 \label{figure_cRF_100}
 \end{figure}

\subsection{Emergent cognitive capabilities}

The simulation results for the $5\times5$ 
bars problem presented in Fig.~\ref{figure_cRP_graph}
may be generalized to larger systems. For
comparison we discuss now the results for
the $10\times10$ bars problem, for which there
are ten horizontal and ten vertical elementary
bars. For the dHan network we used a regular
40-site star with 20 cliques, a straightforward
generalization of the regular star illustrated in
Fig.~\ref{figure_networks_cliques}.
We used otherwise exactly the same set of
parameters as previously for the $5\times5$
bars problem, in particular also the same number
of input training patterns. No optimization of
parameters has been performed.
The respective responses $R(\alpha,\beta)$ and
receptive fields $F(\alpha,j)$ (compare
Eqs.~(\ref{eq_clique_rec_patt}) and 
(\ref{eq_clique_recp_fields})) are presented
in Fig.~\ref{figure_cRP_100} and \ref{figure_cRF_100}.

The probability for any of the 20 bars to occur
in a given input pattern, like the ones for the
$5\times5$ bars problem illustrated in
Fig.~\ref{figure_bars_pattern}, is $p=0.1$
and any individual $10\times10$ input patterns
contains on the average $\approx2.3$ bars
superposed non-linearly. The separation of
the 20 statistically independent components
in the input data stream is therefore a
non-trivial task. The results presented
in Fig.~\ref{figure_cRP_100} indicate that
the system performs the source separation surprisingly 
well, but not perfectly. The respective receptive 
fields, shown in Fig.~\ref{figure_cRF_100}, are 
only in part self-evident. This is, again, due
to the competitive nature of the unsupervised
and local learning process, which has the
task to optimize the input rates to the
dHan layer and not to maximize the 
signal-to-noise ratio. We note in this
context that the system contains no
prior knowledge about the nature
and statistics of the input signals.

In fact, the system has not been constructed 
in the first place to tackle the non-linear independent
component task. The setup used here has been 
motivated by two simple guiding principles,
the occurrence of self-sustained internal
neural activity and the principle of
competitive neural dynamics. These principles
have been used in our study to examine the interplay
of the self-sustained internal neural 
dynamics with the inflow of external information 
via a sensory data stream. One can therefore 
interpret, to a certain extend, the capability 
of the system to perform a non-linear
independent component analysis as
an example of an `emergent cognitive
capability'. This information processing
capability emerges from general construction 
principles and does not result from the
implementation of a specific neural algorithm.

\section{DISCUSSION AND CHALLENGES}

\subsection{Discussion}

A standard approach in the field of neural networks
is to optimize the design of a network such that
a given cognitive or computational task can be
tackled efficiently. This strategy has been
very successful in the past with respect to
technical applications like handwriting
recognition \cite{dreyfus05} and regarding the modeling
of initial feed-forward sensory information processing
in cortical areas like the primary optical 
cortex \cite{arbib02}. Task-driven network design 
standardly results in input-driven neural networks,
with cognitive computation coming to a standstill 
in the absence of sensory inputs. 

Real-world cognitive systems like the human brain are however 
driven by their own internal dynamics and it constitutes
a challenge to present and to future research in the field
of neural networks to combine models of this
self-sustained brain activity with the processing
of sensory data. This challenge regards especially
recurrent neural networks, since recurrency is
an essential ingredient for the occurrence of
spontaneous internal neural activities.

In this work we studied the interplay of self-generated
neural states, the time-series of winning coalitions, 
with the sensory input for the purpose of unsupervised 
feature extraction. We proposed learning to be autonomously 
activated during the transition from one winning 
coalition to the subsequent one.

This general principle may be implemented algorithmically
in various fashions. Here we used a generalized neural
net (dHan - dense homogeneous associative net) for
the autonomous generation of a time series of associatively
connected winning coalitions and controlled the 
unsupervised extraction of input-features by an autonomously
generated diffusive learning signal.

We tested the algorithm for the bars problem and
found good and fast learning and that the initially 
semantically void transient states acquired, 
through interaction with the data input stream,
a semantic significance. Further preliminary results 
indicate that the learning algorithm retains functionality 
under a wide range of conditions and for various sets 
of parameters. We plan to extend the simulations 
to various forms of temporal inputs, especially to 
quasi-continuous input and to natural scene analysis,
and to study the embedding of the here proposed concept
within the framework of a full-fledged and
autonomously active cognitive system.

\subsection{The overall perspective}

There is a growing research effort trying 
to develop universal operating principles for
biologically inspired cognitive systems, the rational
being, that the number of genes in the human genome
is by far too small for the detailed encoding of 
the fast array of neural algorithms the brain is
capable off. There is therefore a growing consensus,
that universal operating principles may be potentially
of key importance also for synthetic cognitive and
complex systems \cite{organic04,organic08}. The present
work is motivated by this line of approach.

Universal operating principles for a cognitive system
remain functionally operative for a wide range of
environmental conditions. Examples are, universal
time prediction tasks for the unsupervised extraction
of abstract concepts and intrinsic generalized grammars
from the sensory data input stream
\cite{elman90,elman04,grosBook08} and the optimization
of complexity and information theoretical measures 
for closed-loop sensorimotor behavioral studies of
simulated robots \cite{seth2004,sporns2006,ay2008}.
The present study is motivated by a similar line
of thinking, investigating the consequences of a
self-sustained internal neural activity in recurrent
networks, being based on the notion of transient-state
and competitive neural dynamics. The long-term goal
of an autonomous cognitive system is pursued in this
approach via a modular approach, with each module
being based on one of the above mentioned general
architectural and operational principles.



\addtolength{\textheight}{-3cm}   

\end{document}